\documentclass[conference]{IEEEtran}
\ifCLASSINFOpdf
\else
\fi
\usepackage{algorithm}
\usepackage{algorithmic}
\usepackage{graphicx}

\begin{document}
\title{An Efficient Multiplication Algorithm Using Nikhilam Method}
\author{\IEEEauthorblockN{Shri Prakash Dwivedi}
\IEEEauthorblockA{Email: shriprakashdwivedi@gbpuat-tech.ac.in}}

\maketitle

\begin{abstract}
Multiplication is one of the most important operation in computer arithmetic. Many integer operations such as squaring, division and computing reciprocal require same order of time as multiplication whereas some other operations such as computing GCD and residue operation require at most a factor of $\log n$ time more than multiplication. We propose an integer multiplication algorithm using Nikhilam method of Vedic mathematics which can be used to multiply two binary numbers efficiently.
\end{abstract}


\IEEEpeerreviewmaketitle

\section{Introduction}
The classical method of adding two integers of $n$-bits takes $O(n)$ bit operations but the classical method of multiplying them takes $O(n^{2})$ bit operations. Complexity of addition is optimal in number of bit operations, whereas optimal multiplication algorithm for integers is an open problem. The classical approach to multiply two $n$-bit integers requires $O(n^{2})$ bit operations. Karatsuba multiplication [9] uses divide-and-conquer technique to multiply two $n$-bit integers in $O(n^{\log 3})$ bit operations(logarithms are to the base 2 unless otherwise specified) by replacing some of the multiplication by less costly addition and subtraction. Toom-Cook algorithm further improves the above bound [3]. Toom-Cook method is the generalization of Karatsuba method which split each number to be multiplied in multiple parts [12]. Given two large integers, Toom-Cook splits up multiplicand and multiplier into $k$ smaller parts each of length $l$, and performs operations on the parts. As $k$ grows, one may combine many of the multiplication sub-operations, thus reducing the overall complexity of the algorithm. For $k=3$ Toom-Cook reduces 9 multiplication to 5, with 
asymptotic complexity of $O(n^{(\log 5)/(\log 3)})$. Schonhage-Strassen integer multiplication algorithm [10] uses Fast Fourier Transform (FFT) by selecting the principal roots of unity as evaluation point to perform multiplication in $O(n \log n \log\log n)$ bit operations. FFT method employs a divide-and-conquer strategy by taking advantage of the special properties of the complex root of unity to perform multiplication of two polynomials in only $O(n \log n)$ arithmetic complexity [4]. In fact many multiplication algorithms can viewed as schemes for the evaluation of polynomials then multiplication of their values and followed by interpolation. Currently, the asymptotically fastest algorithm for multiplication of two $n$-bit integers is by Furer [7] which runs in $O(n \log n 2^{O(\log^* n)})$. Where $\log^* n$ is iterated logarithm function [5] defined as: $$\log^* n=\min\{i\geq 0 : \log^{(i)}n \leq 1\} .$$
Furer algorithm uses arithmetic over complex number. Same asymptotic bound can also be achieved using modular arithmetic [6].\\
\indent The Schonhage-Strassen algorithm and Furer algorithm are asymptotically fast but they are suited for extremely large numbers. Furer algorithm although asymptotically fastest, only achieves an advantage for astronomically large values and as such it is currently not used in practice. The crossover points between these algorithms are usually very high when the algorithms are implemented [8].
For small inputs even Karatsuba algorithm runs slower than the classical multiplication algorithm because of recursion overhead. In this paper we use Nikhilam sutra or method from vedic mathematics [11] to perform efficient multiplication for small inputs. Nikhilam sutra performs large multiplication by converting it to small multiplication along with some addition and shifting operations. \\
\indent This paper is organized as follows. Section 2 describes background and motivation. Section 3 presents our proposed work, the multiplication algorithm and its features. In section 4, we present some applications. Finally, section 5 contains conclusion.

\section{Background and Motivation}
The simplest method to multiply two $n$-digit integers is using classical or long multiplication method which requires
$O(n^{2})$ multiplication operations. Whereas to add or subtract two $n$-digit integers using traditional method requires at most $n$ number of addition or subtraction which is optimal in terms of number of addition/subtraction operation performed. To improve the $O(n^{2})$ bound of multiplication, several algorithms have been discovered.  The simplest one is Karatsuba algorithm which is based on divide-and-conquer paradigm [1]. Karatsuba algorithm is based on the fact that two-digit multiplication can be done with only three instead of four multiplications required by standard method. Suppose we want to multiply two $2$-digit decimal numbers $a_1a_2 * b_1b_2$ :
\begin{enumerate}
 \item[1.] Compute $A = a_1*b_1$
 \item[2.] Compute $B = a_2*b_2$
 \item[3.] Compute $C = (a_1 + a_2)*(b_1 + b_2)$
 \item[4.] Compute $D = C-A-B$, here $D$ is equal to $a_1*b_2 + a_2*b_1$
 \item[5.] Result  $100*A + 10*D +B$
\end{enumerate}
For large number of digits we can apply this method recursively by splitting the multiplicand and multiplier in two parts. The complexity of this method is $O(n^{(\log 3)/(\log 2)})$. Since multiplication operation is costly as compared to addition and shift, some constant number of addition and shift operations can be safely ignored. In this paper we assume that multiplicand and multiplier are having equal number of digits.\\

For example suppose we want to multiply $95 * 96$. The standard method of long multiplication requires 4 one-digit multiplication along with some addition and shift. Using Karatsuba algorithm we can compute it as follows:
\begin{enumerate}
 \item[1.] Compute $A = 9*9$
 \item[2.] Compute $B = 5*6$
 \item[3.] Compute $C = (9 + 5)*(9 + 6)$; $C= 14*15$
 \item[4.] Compute $D = C-A-B= 210-81-30=99$ 
 \item[5.] Result  $100*81 + 10*99 +30= 9120$
\end{enumerate}
A schematic view of above multiplication is shown in Fig. 1. The three multiplication operations are enclosed in ellipse.  In fact the total number of $1$-digit multiplication required in above example is 5. Note that to compute $14*15$ in step $3$ requires, three $1$-digit multiplication by applying this method one's more. 
% FIGURE 1%
\begin{figure}[ht]
\begin{center}
\includegraphics[width=7cm, height=4cm]{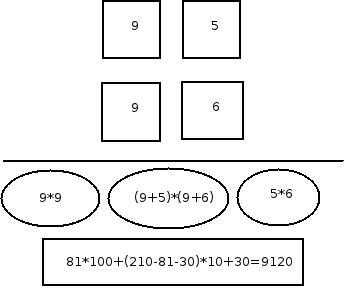}
\end{center}
\caption{Multiplication of integers $(95*96)$ using Karatsuba method}
\label{graph1}
\end{figure}

\indent Nikhilam Sutra is one of the 16 sutras of Vedic mathematics [11]. It can be used to convert large-digits multiplication to small-digits multiplication with the help of few extra add, subtract and shift operations. In some cases two-digit multiplication can be performed using only $1$ one-digit multiplication instead of $3$ one-digit multiplication as required by Karatsuba algorithm.
Suppose we have to perform same multiplication $95 * 96$ using this method. We can use the Nikhilam sutra as follows:
\begin{enumerate}
 \item[1.] Compute $A = 100-95$; Subtract the multiplicand from nearest base 
 \item[2.] Compute $B = 100-96$; Subtract the multiplier from the same base
 \item[3.] Compute $C = B*A=5*4=20 $
 \item[4.] Compute $D = 95-4=96-5=91$ 
 \item[5.] Result  $100*D + C = 9120$
\end{enumerate}

In the Fig. 2, we can see that there is only one multiplication operation involved.
% FIGURE 2%
\begin{figure}[ht]
\begin{center}
\includegraphics[width=7cm, height=4cm]{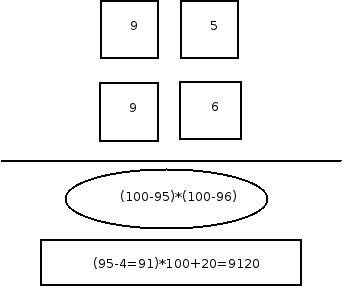}
\end{center}
\caption{Multiplication of integers $(95*96)$ using Nikhilam method}
\label{graph1}
\end{figure}

Above multiplication is also shown in Table 1. In this multiplication we have used $1$ multiplication, $1$ addition, $3$ subtraction and $1$ shift operation. This particular multiplication is more efficient than both standard multiplication and Karatsuba method. Suppose multiplicand is $m=x-a$ and multiplier is $n=x-b$ where $x$ is nearest base. We have: 
$$ m*n=(x-a)*(x-b)=x(x-a-b)+ab $$ The general scheme of multiplication $m*n$ is shown in Table 2.
% TABLE 1%
\begin{table}[ht]
\renewcommand{\arraystretch}{1.3}
\caption{Multiplication of $95*96$}
\label{table 1}
\begin{center}
\begin{tabular}{|c| c| c| } 
\hline
 & Integer & Base Difference    \\ [1ex] \hline\hline 
Multiplicand & 95 & (100-95)=5   \\ \hline
Multiplier & 96 & (100-96)=4  \\ \hline
 & (95-4)=91 & (5*4)=20  \\ \hline
Result & 9120 &   \\ [1ex] \hline

\end{tabular} 
\end{center}
\end{table} 

% TABLE 2%
\begin{table}[ht]
\renewcommand{\arraystretch}{1.3}
\caption{Multiplication of $m*n$}
\label{table 2}
\begin{center}
\begin{tabular}{|c| c| c| } 
\hline
 & Integer & Base Difference    \\ [1ex] \hline\hline 
Multiplicand & $m$ & $x-(x-a)$   \\ \hline
Multiplier & $n$ & $x-(x-b)$  \\ \hline
 & $(x-a-b)$ & $ab$  \\ \hline
Result & $x(x-a-b)+ab$ &   \\ [1ex] \hline

\end{tabular} 
\end{center}
\end{table} 

This scheme can also be utilized, when the multiplicand and multiplier are just above the certain power of base(which is $10$ in this case). Suppose we have to multiply $105*106$. We can proceed as follows:
\begin{enumerate}
 \item[1.] Compute $A = 105-100$; Subtract the multiplicand from nearest base 
 \item[2.] Compute $B = 106-100$; Subtract the multiplier from the same base
 \item[3.] Compute $C = B*A=5*6=30 $
 \item[4.] Compute $D = 105+6=106+5=111$ 
 \item[5.] Result  $100*D + C = 11130$
\end{enumerate}

Above multiplication is shown in Table 3. This $3$-digit multiplication is performed in just single $1$ digit multiplication, whereas the standard method will take $9$ multiplication and Karatsuba can compute in $4$ multiplication operations.

% TABLE 3%
\begin{table}[ht]
\renewcommand{\arraystretch}{1.3}
\caption{Multiplication of $105*106$}
\label{table 3}
\begin{center}
\begin{tabular}{|c| c| c| } 
\hline
 & Integer & Base Difference    \\ [1ex] \hline\hline 
Multiplicand & 105 & (105-100)=5   \\ \hline
Multiplier & 106 & (106-100)=6  \\ \hline
 & (105+6)=111 & (5*6)=30  \\ \hline
Result & 11130 &   \\ [1ex] \hline

\end{tabular} 
\end{center}
\end{table} 

The principle behind this is as follows: 
Let the multiplicand be $m=x+a$ and multiplier be $n=x+b$ where $x$ is the nearest base, then 
$$ m*n=(x+a)*(x+b)=x(x+a+b)+ab $$
Details are given in Table 4.

% TABLE 4%
\begin{table}[ht]
\renewcommand{\arraystretch}{1.3}
\caption{Multiplication of $m*n$}
\label{table 4}
\begin{center}
\begin{tabular}{|c| c| c| } 
\hline
 & Integer & Base Difference    \\ [1ex] \hline\hline 
Multiplicand & $m$ & $(x+a)-x$   \\ \hline
Multiplier & $n$ & $(x+b)-x$  \\ \hline
 & $(x+a+b)$ & $ab$  \\ \hline
Result & $x(x+a+b)+ab$ &   \\ [1ex] \hline

\end{tabular} 
\end{center}
\end{table}

\section{Proposed Work}
\subsection{Binary Multiplication}
We can perform binary digit multiplication using Nikhilam sutra by converting $n$-bit multiplication to $(n-1)$-bit multiplication and some additional add/subtract and shift operation. We can apply this conversion repeatedly until we get trivial multiplicand/multiplier or $1$-bit multiplication. We can also put some threshold limit $m$ where $1<m<n$ up to which we would like to do this conversion.\\

$2$-bit multiplication can be performed using single $1$-bit multiplication. For example if we have to multiply $11*11$. Here multiplicand $M=11$, and multiplier $N=11$. We can proceed as follows:
\begin{enumerate}
 \item[1.] Compute $A = 11-10$; Subtract the multiplicand from nearest base 
 \item[2.] Compute $B = 11-10$; Subtract the multiplier from the same base
 \item[3.] Compute $C = B*A=1*1=1 $
 \item[4.] Compute $D = M+B=N+A=11+1=100$ 
 \item[5.] Result  $10*D + C = 1001$
\end{enumerate}
The only multiplication required in this computation(Table 5) is for $C$ in step 3.
% TABLE 5%
\begin{table}[ht]
\renewcommand{\arraystretch}{1.3}
\caption{Binary Multiplication of $11*11$}
\label{table 5}
\begin{center}
\begin{tabular}{|c| c| c| } 
\hline
 & Bits & Base Difference    \\ [1ex] \hline\hline 
Multiplicand & 11 & (11-10)=1   \\ \hline
Multiplier & 11 & (11-10)=1  \\ \hline
 & (11+1)=100 & (1*1)=1  \\ \hline
Result & 1001 &   \\ [1ex] \hline

\end{tabular} 
\end{center}
\end{table} 

\indent For $3$-bit multiplication consider the example of $101*110$
\begin{enumerate}
 \item[1.] Compute $A = 101-100$; Subtract the multiplicand from nearest base 
 \item[2.] Compute $B = 110-100$; Subtract the multiplier from the same base
 \item[3.] Compute $C = 10*1=10 $
 \item[4.] Compute $D = M+B=N+A=101+10=111$ 
 \item[5.] Result  $100*D + C = 11110$
\end{enumerate}
In this computation also two $1$-bit multiplication is performed. While in case of standard multiplication $9$ multiplication is required, and Karatsuba algorithm use $4$ multiplication.
% TABLE 6%
\begin{table}[ht]
\renewcommand{\arraystretch}{1.3}
\caption{$3$-bit Binary Multiplication of $101*110$}
\label{table 6}
\begin{center}
\begin{tabular}{|c| c| c| } 
\hline
 & Bits & Base Difference    \\ [1ex] \hline\hline 
Multiplicand & 101 & (101-100)=1   \\ \hline
Multiplier & 110 & (110-100)=10  \\ \hline
 & (101+10)=111 & (1*10)=10  \\ \hline
Result & 11110 &   \\ [1ex] \hline

\end{tabular} 
\end{center}
\end{table} 

Let us consider $4$-bit multiplication of $1111*1111$. In this case multiplicand(M) and multiplier(N) are equal i.e. $M=N=1111$. For multiplication we can proceed as follows:
\begin{enumerate}
 \item[1.]  Compute $A = 1111-1000$; Subtract the multiplicand from nearest base 
 \item[2.]  Compute $B = 1111-1000$; Subtract the multiplier from the same base
 \item[3.]  Compute $C = A-100=11 $
 \item[4.]  Compute $D = B-100=11 $ 
 \item[5.]  Compute $E = C-10=1$
 \item[6.]  Compute $F = D-10=1$
 \item[7.]  Compute $G = E*F=1$
 \item[8.]  Compute $H = (C+F)*10 + G=1001$
 \item[9.]  Compute $I = (A+D)*100 + H=110001$
 \item[10.] Result  $J = (M+B)*1000 + I=11100001$
\end{enumerate}
In this computation the only multiplication involved is in step 7 and apart from that all other steps are having either add/subtract or shift operation. Details of this multiplication are shown in Table 7.

% TABLE 7%
\begin{table*}[ht]
\renewcommand{\arraystretch}{1.3}
\caption{$4$-bit Binary Multiplication of $1111*1111$}
\label{table 7}
\begin{center}
\begin{tabular}{|c| c| c| c| c|} 
\hline
 & Bits & Base Difference  & Next Difference & Next Difference  \\ [1ex] \hline\hline 
Multiplicand & 1111 & (1111-1000)=111 &(111-100)=11 & (11-10)=1  \\ \hline
Multiplier & 1111 & (1111-1000)=111 &(111-100)=11 & (11-10)=1  \\ \hline
 &  &  &  & $1*1=1$  \\ \hline
 &  &  & $(11+1)*10+1=1001$ & \\ \hline
 &  & $(111+11)*100 + 1001=110001$ &  & \\ \hline
 & $(1111+111)*1000 + 110001$ &  &  & \\ \hline
Result & 11100001 &  &  &   \\ [1ex] \hline

\end{tabular} 
\end{center}
\end{table*}

From these examples we can easily recognize that the computation is simple when multiplicand and multiplier both are same.

\subsection{Nikhilam Algorithm}
 Nikhilam sutra seems to have special advantage when the multiplicand and multiplier are same. If both multiplicand and multiplier are equal then multiplication operation is known as squaring. Squaring is considered as special case of multiplication. First we write the NikhilamSquaring algorithm which can be used to compute square of a binary integer. NikhilamSquaring algorithm is in turn called by NikhilamMultiplication to perform the multiplication. Multiplication and squaring related to each other by the following well known formula: $$x*y=\frac{(x+y)^2 - (x-y)^2}{4} $$

\indent Description of NikhilamSquaring is given in Algorithm 1. It takes input $A$ as a binary number of $n$-bits and produces square of $A$ as its output which can be up to $2n$-bits. First while-loop is used for the computation of the forward direction subtraction operations. Two counters $i$ and $j$ are used to keep track of processed input and proper base subtraction respectively. If-loop is used to check whether the corresponding bit is $0$ or $1$. If base power to be subtracted is more than the number itself then else part of if-loop is executed. This happens to be only when the first bit of the number is $0$. Least significant bits multiplication is assigned in $B_1$. Second while-loop is used for the computation of reverse direction for shifting and addition operations. Again If-loop is used to check whether two consecutive $A_j$ values are same, if it is so next value of $B_i$ is unchanged, otherwise its value is updated.   

\begin{algorithm}
\caption{\bf :  NikhilamSquaring $(A)$}
\begin{algorithmic}
\STATE \textbf{INPUT}: $A=\sum_{0}^{n-1} a_i x^i$
\STATE \textbf{OUTPUT}: $B=A*A=\sum_{0}^{2n-1} b_k x^k $
   \STATE $A_1 \leftarrow A$ 
   \STATE $i\leftarrow 2, j\leftarrow n-1$
   \WHILE {$(i \leq n$ \textbf{and} $j \geq 1 )$ } 
   {
   \IF {$(A_i > 2^j)$}
   \STATE $A_i \leftarrow A_{i-1} - 2^j $ \\
   \ELSE \STATE $A_i \leftarrow A_{i-1}$
   \ENDIF
   \STATE $i\leftarrow i+1, j\leftarrow j-1$
   }
   \ENDWHILE
   \STATE $B_1 = A_n * A_n$
   \STATE $i\leftarrow 2, j\leftarrow n-1$
   \WHILE {$( i \leq n$  \textbf{and} $j \geq 1 )$ }
   {
   \IF {$A_j \neq A_{j+1}$}
   \STATE $B_i \leftarrow B_{i-1} + (A_j+ A_{j+1})2^{i-1} $ \\
   \ELSE \STATE $B_i \leftarrow B_{i-1} $
   \ENDIF
   \STATE $i\leftarrow i+1, j\leftarrow j-1$
   }
   \ENDWHILE

   \RETURN $B \leftarrow B_n$
\end{algorithmic}
\end{algorithm}

Execution of NikhilamSquaring algorithm for $6$-bit input $A=101010 $ is shown in Table 8. We have $A_1=101010$, $A_2=(101010-100000)=01010$, $A_3=1010$ (since $01010 > 10000$), $A_4=(1010-1000)=010$, $A_5=10$, $A_6=0$, $B_1 =0*0=0$, $B_2 = (10+0)*10+0=100 $, $B_3=100$ (since $A_4 = A_5$), $B_4 = (1010+010)*1000+100=1100100$, $B_5 =1100100$,  $B_6 = (101010+1010)100000 + 1100100= 11011100100 $.

% % TABLE 8%
\begin{table*}[ht]
\renewcommand{\arraystretch}{1.3}
\caption{$6$-bit Binary Multiplication of $101010*101010$}
\label{table 8}
\begin{center}
\begin{tabular}{|c| c| c| c| c| c| c|} 
\hline
 & Binary Digits & Base Difference  & Next Difference & Next Difference & Next Difference & Next Difference \\ [1ex] \hline\hline 
Multiplicand & $ A_1=101010$ & $A_2=01010$ & $A_3=1010$ & $A_4=010$ & $A_5=10$ & $A_6=0$ \\ \hline
Multiplier & $ A_1=101010$ & $A_2=01010$ & $A_3=1010$ & $A_4=010$ & $A_5=10$ & $A_6=0$  \\ \hline
$B_1 = A_6 * A_6$ &  &  &  &  &  & $B_1 =0*0=0$  \\ \hline
$B_2$ &  &  &  &  & $B_2 =100$ & \\ \hline
$B_3$ &  &  &  &$B_3 =100$ &  & \\ \hline
$B_4$ &  &  &$B_4 =1100100$ &  &  & \\ \hline
$B_5$ &  &  $B_5 =1100100$ &  &  &  & \\ \hline
Result & $B_6=11011100100$ &  &  &  &  & \\ [1ex] \hline

\end{tabular} 
\end{center}
\end{table*} 

Correctness of the NikhilamSquaring can be easily established using induction on the bit length of the input $A$. Note that in each column of Table 8, the partial result is, in fact multiplication of corresponding multiplicand and multiplier in that column.\\

\textbf{Theorem 1.} \textit{NikhilamSquaring algorithm computes square of the input $A$.}\\
 
\textbf{Proof}:  We prove this using induction on the bit length $n$ of $A$. For $n=1$, number $A$ has only one bit and $A^2=1$ when $A$ is $1$ and $A^2=0$ when $A$ is $0$. Therefore it works for $n=1$. Assume it is true for $k-1$. Now we show it for $k$. Assume $A_{k-1} = x$, therefore $A_{k-1}^2=x^2 $. \\
 Case 1: If $k$th bit is $1$ then $A_k=2x+1$, and the processing of $k$th step is $$((2x+1)+x)*2^k + x^2 $$ also we have $$(2x+1)-2^k=x \Rightarrow 2^k = x+1 $$ and therefore
 $$((2x+1)+x)*2^k + x^2 = ((2x+1)+x)(x+1) + x^2$$  $$ \Rightarrow (3x+1)(x+1) +x^2 = 4x^2 + 4x+ 1 = (2x+1)^2 $$
 and hence $A_{k}^2=(2x+1)^2 $, and the statement of theorem follows.\\
 Case 2: If $k$th bit is $0$, then the statement is trivial and $ A_k = A_{k-1} = x^2$, and theorem is proved.\\

\indent NikhilamMultiplication is described in Algorithm 2. It takes two binary numbers $A$ and $B$ as input and compute their multiplication$(C)$ as output. It performs two calls to NikhilamSquaring algorithm and using that, outputs the desired result. NikhilamMultiplication uses only $1$ multiplication, $1$ division and $O(c*n)$ add/subtract and shift operations for some constant $c$. To show the correctness of NikhilamMultiplication is trivial given the correctness of NikhilamSquaring.\\

\begin{algorithm}
\caption{\bf :  NikhilamMultiplication$(A, B)$}
\begin{algorithmic}
\STATE \textbf{INPUT}: $A=\sum_{0}^{n-1} a_i x^i, B=\sum_{0}^{n-1} b_j x^j$
\STATE \textbf{OUTPUT}: $C=AB=\sum_{0}^{2n-1} c_k x^k $
\STATE $D_1 \leftarrow$ \textbf{NikhilamSquaring} $(A+B) $
\STATE $D_2 \leftarrow$ \textbf{NikhilamSquaring} $(A-B) $
\RETURN $C \leftarrow (D_1 - D_2)/4 $
\end{algorithmic}
\end{algorithm}

\textbf{Theorem 2.} \textit{NikhilamMultiplication algorithm computes the product of $A$ and $B$.}\\

\textbf{Proof}: The statement of the above theorem follows from Theorem 1 and the fact that :
 $$ A*B = \frac{(A+B)^2 - (A-B)^2}{4} .$$  \\

\section{Applications}
Since, asymptotically fast multiplication algorithms like Schonhage-Strassen and Furer algorithms are only useful for extremely large numbers, for small to medium size numbers we can apply Nikhilam multiplication. We can use Nikhilam multiplication even in conjunction with some other fast algorithm like Karatsuba. If $n_0$ is the threshold between classical multiplication and Karatsuba algorithm then up to threshold limit $n_0$ we can use Nikhilam multiply and beyond that limit we can use Karatsuba multiply. We can write the Karatsuba algorithm as given in [2], with the only modification that if $n<n_0$ NikhilamMultiplication is called. The corresponding pseudo code is given in Algorithm 3. Karatsuba multiplication has relatively small threshold with the classical multiplication. The optimal threshold for Karatsuba algorithm can vary from about ten to hundred words. NikhilamMultiplication can also be used as a stand alone multiplication algorithm.\\
\indent Further optimization to NikhilamMultiplication is also possible. Least significant $0$'s can be truncated from input numbers, and in the end corresponding modification can be done in single operation. We can process the multiplicand and multiplier for consecutive $0$'s to skip some of the steps of the algorithm. The proposed algorithm is particularly efficient because multiplication operation is least involved in it.

\begin{algorithm}
\caption{\bf :  KaratsubaMultiplication$(A, B)$}
\begin{algorithmic}
\STATE \textbf{INPUT}: $A=\sum_{0}^{n-1} a_i x^i, B=\sum_{0}^{n-1} b_j x^j$
\STATE \textbf{OUTPUT}: $C=AB=\sum_{0}^{2n-1} c_k x^k $
\IF {$(n<n_0)$}
\RETURN \textbf{NikhilamMultiplication} $(A, B)$
\ENDIF
\STATE $K \leftarrow (n/2)$
\STATE $(A_0, B_O) \leftarrow (A,B)$ mod $x^k ,(A_1, B_1) \leftarrow(A, B)$ div $x^k$ 
\STATE $s_A \leftarrow $ sign$(A_0-A_1)$, $s_B \leftarrow $ sign$(B_0-B_1)$
\STATE $C_0 \leftarrow$ \textbf{KaratsubaMultiplication} $(A_0, B_0)$
\STATE $C_1 \leftarrow$ \textbf{KaratsubaMultiplication} $(A_1, B_1)$
\STATE $C_2 \leftarrow$ \textbf{KaratsubaMultiplication} $(|A_0-A_1|, |B_0-B_1|)$
\RETURN $C \leftarrow C_0 +(C_0 + C_1 - s_A s_B C_2)x^k + C_1 x^{2k} $
\end{algorithmic}
\end{algorithm}

\section{Conclusions and Future Work}
In this paper we have explored the possibility of applying the Nikhilam sutra of Vedic mathematics to binary number multiplication. We can take advantage of the fact that this sutra can convert large-digit multiplication to corresponding small digit multiplication. Nikhilam method is particularly efficient when both multiplicand and multiplier are near to some base (radix) power. To take this advantage, we have first performed square operation in NikhilamSquaring and then we have used this to finally compute multiplication.
\indent Future work can be to extend this method to large-digit multiplication and exploit it's properties to perform fast integer multiplications.

\end{document}